# SUPERCONDUCTING FLYWHEEL MODEL FOR ENERGY STORAGE APPLICATIONS

R.V. VIZNICHENKO [1*], A.I. PLYUSHCHAY [1], A.A. KORDYUK [1]

[1] Institute of Metal Physics, 36 Vernadsky Str., 03142 Kiev, Ukraine

*Corresponding author e-mail: vrv@imp.kiev.ua*

**Abstract**

In order to explore the complexity and diversity of the flywheels' dynamics, we have developed the real-physics computer model of a universal mechanical rotor. Due to an arbitrary external force concept, the model can be adjusted to operate identical to the real experimental prototype. Taking the high-speed magnetic rotor on superconducting bearings as the prototype, the law for the energy loss in real high temperature superconducting bearings has been derived. Varying the laws of damping and elasticity in the system, we have found a way to effectively damp the parasitic resonances and minimize the loss of energy storage.

Utilization of the superconducting levitation phenomena in large scale mechanical devices of low energy consumption, like flywheels for energy storage, is an obvious but promising application of high temperature superconductors (HTS) [1]. The idea of the superconducting flywheel or a superconducting bearing in general, is simple: there is no energy loss in the system where rotation is realized coaxially to completely axisymmetric magnetic field. Therefore, the problem seems to converge to purely technological task – to make the system as perfect as possible. However, it is easy to understand that such a 'perfect rotor' encounters a stability problem – as lower the damping as more complicated becomes the rotor's dynamics. This complexity, which appears as mode-entangled parasitic resonances with further evolution in a non-analytic chaotic regime and often leads to a crash of the system, sets technical limits for the flywheels' efficiency [2, 3].

In this paper we explore the complexity and diversity of the flywheels' dynamics by means of the real-physics computer model of a universal mechanical rotor. We study the mechanisms of energy loss as well as parasitic resonances in high-speed magnetic rotor on superconducting bearings and compare results with experimental prototype.

The modeling is based on an accurate calculation of time evolution of space coordinates of a rotating solid body under the action of external forces. In our case the rotor is a cylindrical tube with two magnets on the ends (bearing magnets), which form superconducting bearings and one magnet in the center (driving magnet), which is used to speed-up the rotor by means of driving coil. The magnetic moments of the bearing magnets are parallel to the rotor's symmetry axis, while the moment of the driving magnet is perpendicular to it [4, 6].

Here we study a slightly imperfect rotor which has its magnetic and geometric axes slightly misaligned. Under the driving force of increasing frequency, such a rotor will make low-frequency oscillations which turn into low speed rotation around the magnetic axis. With increase of the speed the axis of the rotation is moving to the geometric axis (the rotor's axis of inertia) [4]. In general, a model where a body is elastically fixed in two arbitrary points appears to be quite applicable to the magnetic rotor on HTS bearings [4]. Therefore, the influence of all external forces can be reduced to three forces – $\mathbf{F_1}$, $\mathbf{F_2}$, $\mathbf{F_{CM}}$, which are applied to two arbitrary points and to the center of mass respectively, and the moment of forces about the mass center $\mathbf{M}$. In general case, all these parameters can depend on time, coordinates and velocities.

To completely describe the dynamics of the above mentioned model it is sufficient to consider the following set of equations:

$$\mathbf{F}_{1,2} = \mathbf{F}_{el}^{1,2}(\delta\mathbf{r}) + \mathbf{F}_{loss}^{1,2}(\mathbf{v},\delta\mathbf{r}) = -k_{1,2}\delta\mathbf{r} + [-\eta_{1,2}\mathbf{v} + \mathbf{f}_{1,2}(\delta\mathbf{r})], \quad (1)$$

$$\mathbf{F}_{CM} = m\mathbf{g} + \mathbf{F}_{ext}(t,\varphi), \quad (2)$$

$$M = -\sigma\omega + M_{ext}(t,\varphi), \quad (3)$$

where $\mathbf{F}_{el}$ and $\mathbf{F}_{loss}$ are elastic force and 'HTS-losses' force, which are applied to points 1 and 2, $k$ and $\eta$ – are elasticity and viscosity parameters, $\mathbf{f}(\delta\mathbf{r})$ is the force which produces the hysteretic loss in HTS [5], $\sigma$ is the viscosity of environment (air), $mg$ is the gravity force, $\omega$ is the rotors' angular velocity, $\mathbf{F}_{ext}$ and $M_{ext}$ are the external force and the moment of the external forces to model the interaction of the driving magnet with driving coils and other external disturbances. As output of the model we measure time dependencies of frequency and energy of the rotor: $\omega(t)$, $W(t)$ and $W(\omega)$.

The most important parameter for the flywheel applications is energy loss, especially in HTS bearings. For the perfect rotor and magnets the rotation will not produce the change of magnetic field around the rotor and no AC magnetic field will be generated. In real case, a small AC magnetic field with amplitude $\mathbf{h}_0(\mathbf{r})$ is usually produced by rotating rotor and this amplitude is constant in steady regime. Then we have AC loss in the nearest to the rotor objects, in particular in HTS bearings. So, we can use the described imperfect rotor as a convenient tool to study mechanisms of AC loss in the HTS in a wide range of frequencies [4, 5].

Hence, it is not easy to determine the function of $\mathbf{h}_0(\mathbf{r})$, which depends on the type of rotors' imperfection and here the modelling of resonances helps to solve the problem. For example, Fig. 1 shows the modelling results of the vertical rotor with a simple non-ideallity type: magnetic and inertia axes are parallel but separated by distance $d \approx 0.03$ cm, for three mechanisms of energy loss: (i) – viscous one ($\mathbf{F}_{loss} = -\eta\mathbf{v}$), (ii) – hysteretic one ($\mathbf{F}_{loss} \propto (\mathbf{v}/|\mathbf{v}|)\delta r^2$), which corresponds to hysteretic loss in the critical state model ($W \propto h_0^3$) and (iii) – dry friction force loss ($\mathbf{F}_{loss} = (\mathbf{v}/|\mathbf{v}|)\cdot$const). The others parameters are $\mathbf{F}_{el} = -k\delta\mathbf{r}$, $k = 7.1\times10^4$ g/s$^2$, the rotor's mass $m = 10$ g, moment of inertia $J = 1.8$ g cm$^2$.

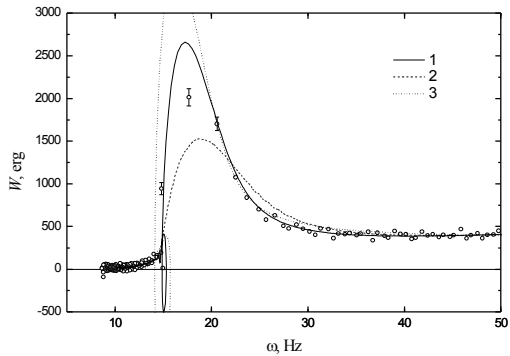

Figure 1. Experimental data (O) and simulation results for three types of energy loss: 1 – viscous, 2 – hysteretic, 3 – dry friction. Magnetic and inertia axes are parallel but separated.

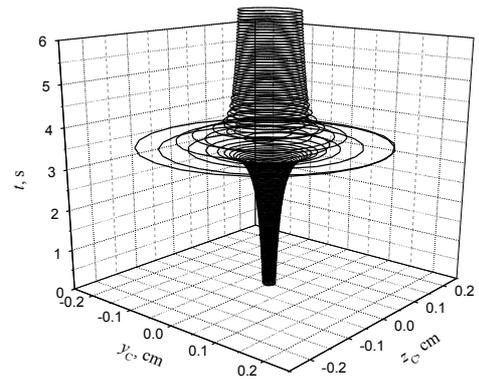

Figure 2. A trajectory of the rotor's center of gravity, the $x$ axis is parallel to the angular velocity vector $\omega$.

The structure of this resonance is presented in Fig. 2, where one can see rotor's center of mass moves around circles, which significantly increase the loss in the system. In the experiment in Fig. 1 the HTS bearings were made from granular superconductors [6] and the best coincidence gives model with viscous a mechanism, that is confirmed by another data [4].

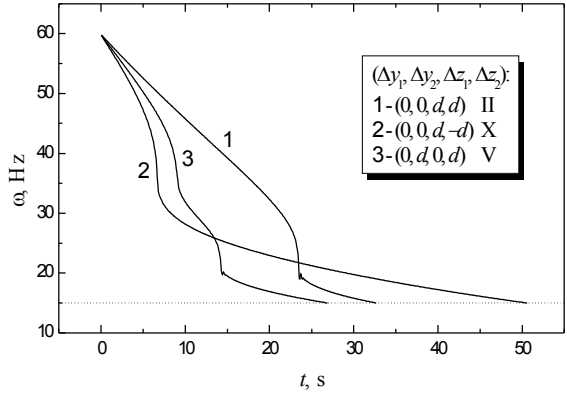
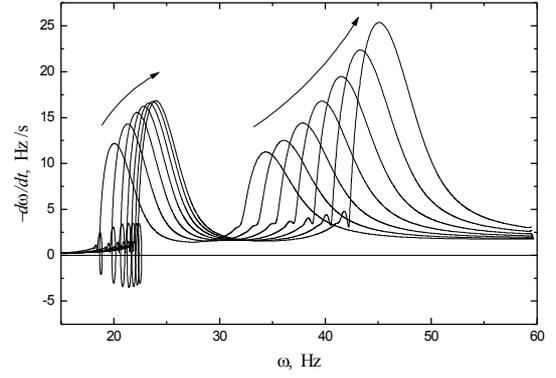

Figure 4. Evolution of resonance maximums of $\dot{\omega}(\omega)$ in case of $\Delta\mathbf{r}_{1,2} = (0, 0, d, d)$ type of non-ideallity. Arrows indicate the increase of $k_2$ from $0.8\,k_1$ to $2.0\,k_1$.

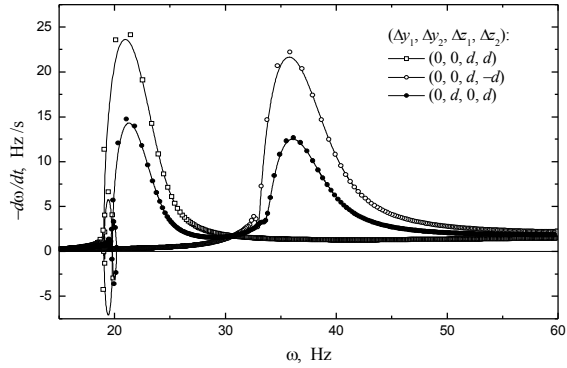
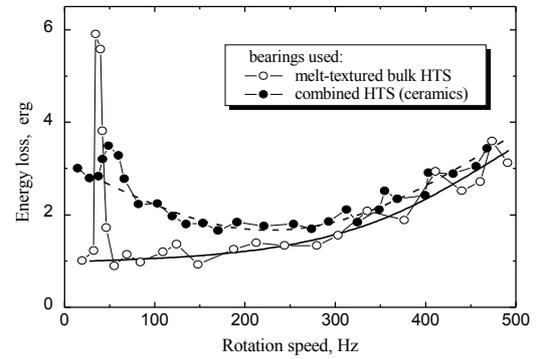

Figure 3. Dependence of $\omega(t)$ and their derivatives for three non-ideallity types with different displacements of the rotor's fixing points from the symmetry axis $\Delta\mathbf{r}_{1,2} = (\Delta y_1, \Delta y_2, \Delta z_1, \Delta z_2)$.

Figure 5. Effect of damping of the parasitic resonance peak by using the combined HTS bearings.

If the geometric and magnetic axes of the rotor are not parallel the dynamics of rotor is more complex. In Fig. 3 we consider three types of imperfection, which are defined by displacement $\Delta\mathbf{r}_{1,2} = (\Delta y_1, \Delta y_2, \Delta z_1, \Delta z_2)$ of the rotor's elastically fixed points with respect to its inertia axis: $1 - (0, 0, d, d)$; $2 - (0, 0, d, -d)$; $3 - (0, d, d, 0)$. Here $d = 0.015$ cm, $k_1 = k_2$, $\eta_1 = \eta_2$ and another parameters are the same. In general cases when $k_1 \neq k_2$, even if the rotor's mass center moves in-plane $\Delta\mathbf{r}_{1,2} = (0, 0, d, d)$, the dependencies of $\dot{\omega}(t)$ and $\dot{\omega}(\omega)$ have two resonance maximums, which shift with change of $k$, see Fig. 4.

Thus the position and size of resonance peaks provide information about rotor's non-ideallity types and possible mechanisms of energy loss. Indeed, the rotor's energy loss per period far from resonances is easily defined as

$$W(\omega) = -\frac{2\pi}{\omega}\frac{dW_0}{dt} = -2\pi J\dot{\omega}(\omega), \qquad (4)$$

where $W_0 = (1/2)J\omega^2$ – the kinetic energy of rotor rotating over its inertia axis, while in resonance region energy of the system looks like the sum of kinetic energy of rotation, energy of mass center motion and potential elastic energy. In case of planar motion of the rotor we have

$$W_0(\omega, v, dr) = \frac{1}{2}J\omega^2 + \frac{1}{2}mv^2 + kdr^2. \qquad (5)$$

Each energy type in (5) has a great oscillation during resonances and the total energy can spread between them leading to complex dynamics of the system.

The knowledge of energy loss structure and its distribution in resonances let us suggest simple approach to reduce effect of parasitic resonances in flywheel systems by using of combined HTS bearings which consist of melt-textured bulk superconductor and ceramic granular superconductor [6]. The former has more large ac loss in low frequency region where the resonances occur, while in high frequencies the loss are negligible and compared with that of in bulk HTS. So, the adding of granular HTS successfully damp parasitic resonances and almost do not influence on the overall flywheel operation, Fig. 5.

In summary, we have modeled the dynamics of the rotor on HTS bearings based on motion simulation of solid body elastically fixed in two points. We have shown the structure of the resonances in a flywheel system is closely related with type of asymmetry of the rotor and magnets as well as with mechanisms of energy loss in superconducting bearings. We have compared the calculated results with ones obtained on the experimental prototype. The way to effectively damp the parasitic resonances and minimize the loss of energy storage by using combined HTS bearings has been suggested.

**References**

[1] J. R. Hull: *Supercond. Sci. Technol.*, **13** (2000) p. R1
[2] A. A. Kordyuk and V. V. Nemoshkalenko: *J. Low Temp. Phys.*, **130** (2003) p. 207
[3] A. A. Kordyuk, V. V. Nemoshkalenko, and H. C. Freyhard: *Adv. Cryo. Eng.*, **43** (1998) p. 727
[4] Kordyuk A.A., Nemoshkalenko V.V.: *IEEE Trans. Appl. Supercond.*, **7** (1997) p. 928
[5] Kordyuk A.A., Nemoshkalenko V.V., Viznichenko R.V., Gawalek W.: *Mat. Sci. Eng. B*, **53** (1998) p. 174
[6] A. A. Kordyuk, V. V. Nemoshkalenko, A. I. Plyushchay, R. V. Viznichenko: *NATO Sci. Ser. E: Applied Sciences*, **356** (1999) p. 583